\documentclass[cite]{PoS}
\usepackage{amsmath}

\newcommand{\lsim}{
\mathrel{\hbox{\rlap{\hbox{\lower4pt\hbox{$\sim$}}}\hbox{$<$}}}}

\newcommand{\gsim}{
\mathrel{\hbox{\rlap{\hbox{\lower4pt\hbox{$\sim$}}}\hbox{$>$}}}}

\title{New Physics Signatures in Kaon Decays}

\ShortTitle{New Physics Signatures in Kaon Decays}

\author{\speaker{Monika Blanke}\\
        CERN Theory Division, CH-1211 Geneva 23, Switzerland\\
        E-mail: \email{monika.blanke@cern.ch}}

\abstract{Kaon physics provides a unique opportunity to identify new flavour and CP violating interactions beyond the Standard Model (SM).
In the SM, implied by the hierarchical structure of the CKM matrix and the GIM mechanism, flavour changing neutral current processes are most strongly suppressed in the kaon sector while the suppression is much less effective in the $B$ meson systems.
Thus their theoretical cleanness makes rare $K$ decays, in particular the $K\to \pi\nu\bar\nu$ system, extremely well suited to look for deviations from their tiny SM values.
Despite the increasingly stringent constraints on new physics from direct search experiments as well as indirect searches in $B$ meson decays, large enhancements of both $K^+\to\pi^+\nu\bar\nu$ and $K_L\to\pi^0\nu\bar\nu$ are still possible, and deviations from the SM could be observed even for a multi-TeV new physics scale.
In addition the correlation between the charged and neutral $K\to\pi\nu\bar\nu$ modes provides insight on the new physics operator structure in $K^0 - \bar K^0$ mixing and its interplay with rare $K$ decays. Useful model-discriminating correlations exist also in the $K_L\to\pi^0\ell^+\ell^-$ system.
Finally the $K\to\ell\nu$ decays provide a clean test of lepton universality and place constraints on new physics complementary to the ones obtained from the searches for charged lepton flavour violating decays.

}

\FullConference{2013 Kaon Physics International Conference,\\
		29 April-1 May 2013\\
		University of Michigan, Ann Arbor, Michigan - USA}

\begin{document}

\section{Introduction}

Kaon physics has played a prominent role in the development of the Standard Model (SM). The observation of the ``strange'' $K$ mesons in cosmic rays  led to the introduction of the three quark model to describe the observed meson and baryon spectra~\cite{GellMann:1964nj}. 
Subsequently in 1970 the charm quark was predicted to explain the observed branching ratio for the decay $K_L\to\mu^+\mu^-$~\cite{Glashow:1970gm}, and was discovered only four years later.
Also the existence of a third generation of quarks was predicted from kaon data: Kobayashi and Maskawa realized that the observed CP violation in the neutral $K$ meson system can be explained within the SM only in the presence of at least three quark flavours that mix with each other~\cite{Kobayashi:1973fv}.

Subsequently the role of kaon physics has shifted to constraining the parameter space of the SM. The most precise determination of the CKM element $|V_{us}|$, the so-called Cabibbo angle~\cite{Cabibbo:1963yz}, is currently obtained from $K$ decays through charged current interactions~\cite{Antonelli:2010yf}. Furthermore the parameter $\varepsilon_K$ measuring CP violation in $K^0-\bar K^0$ mixing, generated at loop level in the SM, provides important information for the determination of the CKM matrix.

With the great success of the  $B$-factories Belle and BaBar confirming the CKM matrix as the dominant source of flavour and CP violation the interest in flavour changing neutral current (FCNC) processes has shifted from a precise determination of the CKM parameters to the search for non-SM contributions to these decays. In order to appreciate the special role played by the $K$ sector, it is instructive to first consider the pattern of effects predicted in the SM. Due to the hierarchical structure of the CKM matrix, together with the GIM suppression~\cite{Glashow:1970gm} of the charm quark contribution, the generic prediction for the size of FCNC transition in the various meson systems is determined by
\begin{equation}
\underbrace{|V_{ts}^* V_{td}|}_{K\text{ system}} \sim 5\cdot 10^{-4} \ll 
\underbrace{|V_{tb}^* V_{td}|}_{B_d\text{ system}} \sim  10^{-2} < 
\underbrace{|V_{tb}^* V_{ts}|}_{B_s\text{ system}} \sim  4\cdot 10^{-2}\,,
\end{equation}
i.\,e.\ FCNC transitions in the kaon sector are most suppressed while the effects in $b\to d$ and $b\to s$ transitions are larger.

The new physics (NP) flavour structure on the other hand does in general not exhibit the CKM hierarchies. Consequently the largest deviations from the SM predictions are to be expected in kaon physics, while the effects in rare $B$ decays are generally smaller. Such a pattern of NP effects can indeed be found e.\,g.\ in the Littlest Higgs model with T-parity (LHT) \cite{Blanke:2006sb,Blanke:2006eb,Blanke:2009am}, in the custodially protected Randall-Sundrum model (RSc) \cite{Blanke:2008zb,Blanke:2008yr,Albrecht:2009xr} or in a general left-right model (LR) \cite{Blanke:2011ry}. Therefore even with the SM-like measurements of the $B_s$ mixing phase and the branching ratio for $B_s\to\mu^+\mu^-$ at LHCb, large NP signatures can still be hoped for in rare kaon decays, such as the $K\to \pi\nu\bar\nu$ system or the $K_L\to\pi^0\ell^+\ell^-$ decays.

\section{\boldmath Lessons from $K^0-\bar K^0$ mixing}
 
Before discussing the possible NP signatures in rare $K$ decays, let us briefly review the lessons we have learned from the study of neutral kaon mixing. In the SM the short-distance contribution to $K^0 - \bar K^0$ mixing is generated first at the one loop level via box diagrams with virtual up-type quarks and $W^\pm$ bosons, and is therefore governed by a single effective operator $(\bar sd)_{V-A}(\bar sd)_{V-A}$. In addition CP conserving quantities are affected by long distance contributions with virtual two or three pion intermediate states which are challenging to predict theoretically. 

CP violating quantities, such as the parameter $\varepsilon_K$ measuring the amount of indirect CP violation in $K_{L,S}\to\pi\pi$ decays, on the other hand are governed by short distance physics only and therefore theoretically much cleaner. In fact with the recent progress determining the bag parameter $\hat B_K$ on the lattice~\cite{lattice}, the uncertainty of the SM prediction is nowadays dominated by parametric uncertainties in particular from the determination of the CKM parameter $|V_{cb}|$. The NNLO  prediction reads~\cite{Brod:2010mj, Brod:2011ty}
\begin{equation}
|\varepsilon_K|_\text{SM} = (1.81\pm 0.28)\cdot10^{-3}
\end{equation}
while experimentally we have~\cite{Beringer:1900zz}
\begin{equation}
|\varepsilon_K|_\text{exp} = (2.228\pm 0.011)\cdot10^{-3}\,.
\end{equation}
Although consistent with each other, these numbers indicate the possibility for a small NP contribution enhancing $\varepsilon_K$ over the SM prediction.

Many popular NP models however yield very large contributions to $K^0 - \bar K^0$ mixing and, in the presence of a generic $\mathcal{O}(1)$ complex phase,  enhance $\varepsilon_K$ by several orders of magnitude over the SM prediction. The absence of a flavour protection mechanism together with the chiral enhancement of the non-SM left-right operator contributions leads to the model-independent bound~\cite{Bona:2007vi}
\begin{equation}
\Lambda_\text{NP}\gsim 10^5\,\text{TeV}\,.
\end{equation}
In turn we can conclude that if NP is present at the TeV scale, it must have a very non-generic flavour structure.

\section{\boldmath Visions for rare $K$ decays}

With the stringent constraints from $K^0 - \bar K^0$ mixing one might naively conclude that there is little hope to discover NP in kaon decays.
However as rare $K$ decays are in general governed by different structures than $\Delta S=2$ physics, it turns out that in many NP models large deviations from the tiny SM predictions can still be hoped for. In addition the correlated study of various rare $K$ decay branching ratios allows to shed light on the operator structure generating $\Delta S=2$ and $\Delta S=1$ transitions.
 
\subsection{\boldmath The $K\to\pi\nu\bar\nu$ system}

The charged and neutral $K\to\pi\nu\bar\nu$ modes play a unique role on the stage of flavour physics. They are governed by a single effective operator $(\bar sd)_V(\bar\nu\nu)_{V-A}$ both in and beyond the SM, so that the effective Hamiltonian can conveniently be written as
\begin{equation}
\mathcal{H}_\text{eff}= 
\frac{G_F}{\sqrt{2}}\frac{\alpha}{2\sin^2\theta_W}
\Big[ V_{cs}^* V_{cd} X_\text{NNL}(x_c) + V_{ts}^* V_{td} |X| e^{i\theta_X} \Big] (\bar sd)_V (\bar\nu\nu)_{V-A}\,.
\end{equation}
Here $X_\text{NNL}(x_c)$ is the charm quark contribution known at the NNLO level~\cite{Buchalla:1993wq,Misiak:1999yg,Buras:2006gb,Brod:2010hi} and relevant only for the CP conserving $K^+\to\pi^+\nu\bar\nu$ mode. The function $X = |X| e^{i\theta_X}$ is a sum of the SM top loop function $X(x_t)$ and the possible NP contribution.

Since the relevant hadronic matrix element can be measured precisely in $K^+\to\pi^0e^+\nu$ decays and the leading isospin breaking corrections are known, the $K\to\pi\nu\bar\nu$ system is essentially free of any non-perturbative uncertainties. The main uncertainties in the SM prediction~\cite{Buchalla:1993wq,Misiak:1999yg,Buras:2006gb,Brod:2010hi}
\begin{eqnarray}
Br(K^+\to\pi^+\nu\bar\nu)_\text{SM} &=& (8.5\pm0.7)\cdot 10^{-11} \,,\\
Br(K_L\to\pi^0\nu\bar\nu)_\text{SM} &=& (2.6\pm0.4) \cdot 10^{-11}
\end{eqnarray}
stem in fact from the uncertainties in the determination of the relevant CKM parameters, in particular $|V_{cb}|$.

On the experimental side we have~\cite{Artamonov:2008qb,Ahn:2009gb}
\begin{eqnarray}
Br(K^+\to\pi^+\nu\bar\nu)_\text{exp} &=& 17.3^{+11.5}_{-10.5}\cdot 10^{-11} \,,\\
Br(K_L\to\pi^0\nu\bar\nu)_\text{exp} &<& 2.6 \cdot 10^{-8} \,.
\end{eqnarray}

\begin{figure}
\center{\includegraphics[width=.45\textwidth]{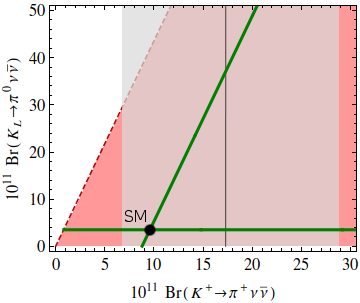}}
\caption{Model-distinguishing correlation between $Br(K^+\to\pi^+\nu\bar\nu)$ and $Br(K_L\to\pi^0\nu\bar\nu)$ \cite{Blanke:2009pq}. See text for details.\label{fig:kpinunu-corr}}
\end{figure}

Through the measurement of both the CP conserving mode $K^+\to\pi^+\nu\bar\nu$ and the CP violating decay $K_L\to\pi^0\nu\bar\nu$ we can determine both the magnitude $|X|$ of the short distance contribution and its NP phase $\theta_X$. Pinning down both parameters could not only leave us with a clear deviation from the SM prediction but in fact, in the presence of such a deviation, could shed some light on the correlation between NP contributions to $\Delta S=2$ and $\Delta S=1$ processes~\cite{Blanke:2009pq}. In models where, like in the SM, only left-handed FCNCs are present and $\Delta S=2$ and $\Delta S=1$ physics is strongly correlated, the stringent constraint from $\varepsilon_K$ on the CP violating phase leaves us with a strong correlation in the $K\to\pi\nu\bar\nu$ system where only two branches of points are allowed, see Fig.\ \ref{fig:kpinunu-corr}. On the other hand if the chirally enhanced left-right operators contribute to $K^0 -\bar K^0$ mixing, the strong correlation with rare $K$ decays is lost and the full $K\to\pi\nu\bar\nu$ plane consistent with the Grossman-Nir bound~\cite{Grossman:1997sk} is possible.

\begin{figure}
\includegraphics[width=.49\textwidth]{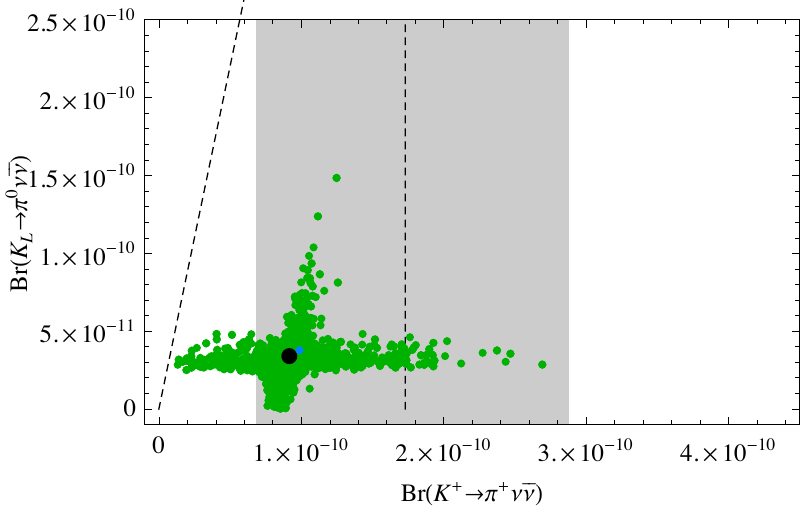}\hfill
\includegraphics[width=.49\textwidth]{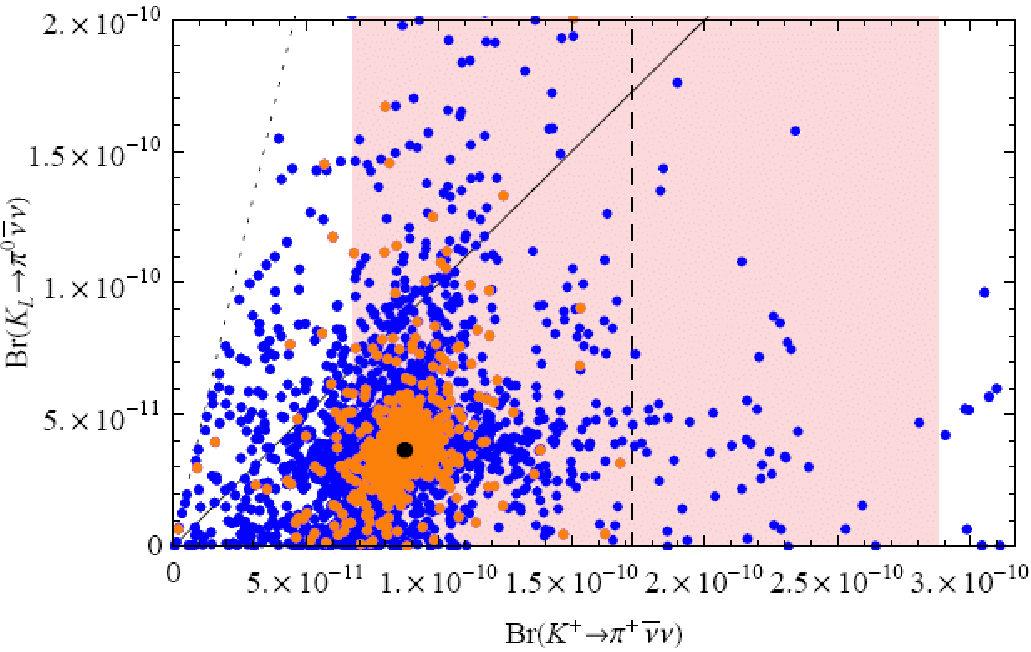}
\caption{Correlation between the branching ratios for $K_L\to\pi^0\nu\bar\nu$ and $K^+\to\pi^+\nu\bar\nu$ in the Littlest Higgs model with T-parity {\it (left)} \cite{Blanke:2009am} and in the custodial Randall-Sundrum model {\it (right)} \cite{Blanke:2008yr}.\label{fig:LHT-RSc}}
\end{figure}

Fig.\ \ref{fig:LHT-RSc} shows the prediction obtained for the $K\to\pi\nu\bar\nu$ decays in two specific NP models, the Littlest Higgs model with T-parity (LHT, left panel)~\cite{Blanke:2008yr} and the Randall-Sundrum model with custodial symmetry (RSc, right panel)~\cite{Blanke:2009am}. Both models have only started to be probed by direct searches at the LHC, and since they naturally predicted rather small effects in $B$ physics observables recently probed by LHCb, large non-SM effects in rare $K$ decays are still possible in both scenarios. In addition we clearly observe the distinguishing power of the correlation discussed model-independently in~\cite{Blanke:2009pq}.
In the LHT model flavour violating interactions, as in the SM, are exclusively left-handed and therefore a strong correlation between $\Delta S=2$ and $\Delta S=1$ transitions exists, resulting in the clear two branch structure in the $K\to\pi\nu\bar\nu$ plane.
On the other hand in the RSc model $\Delta S=2$ transitions are completely dominated by KK gluon exchange inducing the chirally enhanced left-right operators. Consequently no correlation in the $K\to\pi\nu\bar\nu$ plane is visible.

\begin{figure}[h]
\center{\includegraphics[width=.45\textwidth]{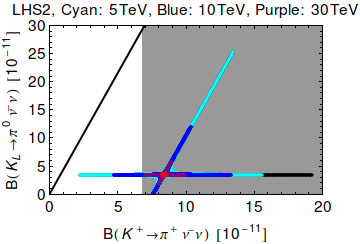}}
\caption{The $K\to\pi\nu\bar\nu$ system in the presence of a $Z'$ gauge boson with exclusively left-handed flavour violating interactions~\cite{Buras:2012jb}.\label{fig:Zprime}}
\end{figure}

The impressive NP discovery reach of rare $K$ decays can be demonstrated by the simple phenomenological assumption of a flavour changing $Z'$ as the lightest NP state~\cite{Buras:2012jb}. From Fig.\ \ref{fig:Zprime} we see that observable effects in the $K\to\pi\nu\bar\nu$ system are possible even for $Z'$ masses as large as $20-30\,\text{TeV}$. Moreover, again we observe the clear two branch structure originating from the pure left-handedness of flavour violating interactions assumed in this particular scenario.

Last but not least let us mention the situation in the minimal supersymmetric SM (MSSM). In that case the main contribution to the $K\to\pi\nu\bar\nu$ decays are coming from wino loops~\cite{Colangelo:1998pm}. Due to the necessity of $SU(2)_L$ breaking interactions the necessary flavour violation is provided by the up squark trilinear couplings. Due to the unique sensitivity of the rare $K$ decays to these couplings, sizeable effects in the $K\to\pi\nu\bar\nu$ decays are not excluded by the recent LHCb data. Similarly the bounds from direct searches at ATLAS and CMS do not place a strong limit due to the slower decoupling of MSSM contributions in $\Delta S=1$ processes with respect to $\Delta S=2$~\cite{Isidori:2006qy}.

\subsection{\boldmath The $K_L\to\pi^0\ell^+\ell^-$ decays}

\begin{figure}
\center{\includegraphics[width=.45\textwidth]{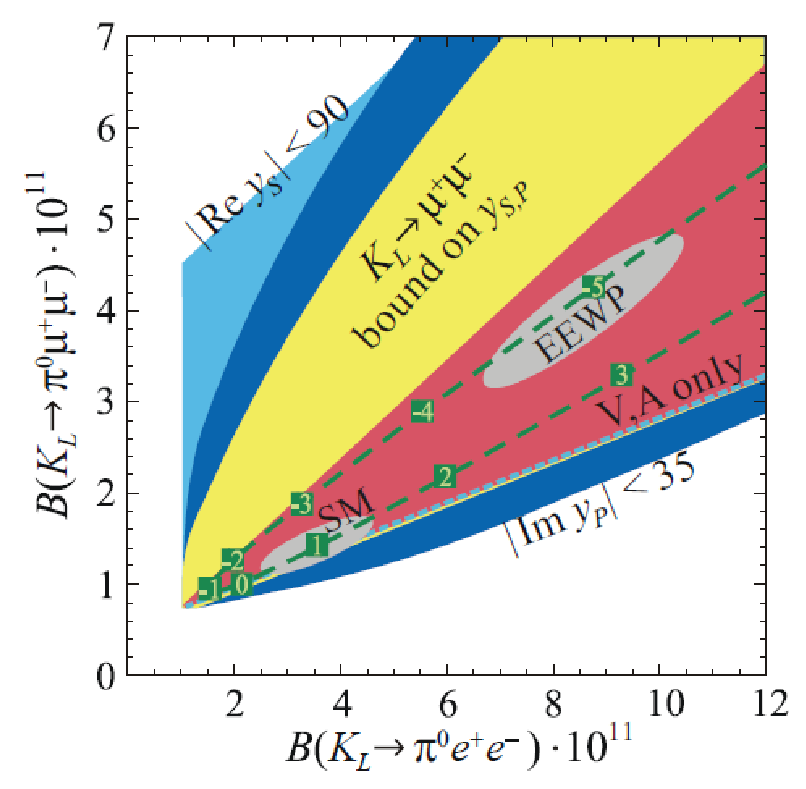}}
\caption{Correlation between $K_L\to\pi^0 e^+e^-$ and $K_L\to\pi^0\mu^+\mu^-$
that allows to test the operator structure of $\Delta S=1$ transitions~\cite{Mescia:2006jd}.\label{fig:Kpll}}
\end{figure}

Additional important information on the NP flavour structure can be obtained from the decays $K_L\to\pi^0 e^+e^-$ and $K_L\to\pi^0\mu^+\mu^-$. While these decays cannot compete with the theoretical cleanness of the $K\to\pi\nu\bar\nu$ system (see \cite{Buchalla:2008jp} for a review), they offer complementary information on the NP operator structure not accessible through the $K\to\pi\nu\bar\nu$ modes. While the $K_L\to\pi^0 e^+e^-$ decay, similarly to the $K\to\pi\nu\bar\nu$ modes, is sensitive only to NP in $Z$ penguins and box diagrams, scalar operator contributions can enter $K_L\to\pi^0\mu^+\mu^-$. The correlation between the two decay rates then serves as a powerful probe of the size of these contributions, as shown in Fig.\ \ref{fig:Kpll}. Similarly the correlation between $K_L\to\pi^0\mu^+\mu^-$ and $K_L\to\pi^0\nu\bar\nu$ can be used~\cite{Blanke:2006eb}.

\subsection{\boldmath A word on $\varepsilon'/\varepsilon$}

\begin{figure}
\center{\includegraphics[width=.45\textwidth]{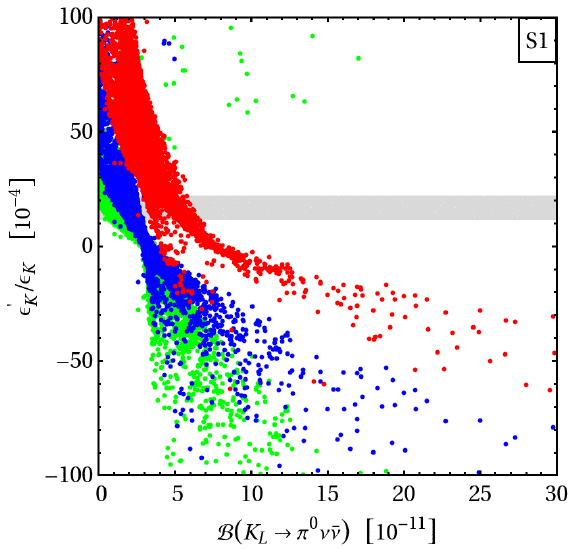}}
\caption{Correlation between $K_L\to\pi^0 \nu\bar\nu$ and $\varepsilon'/\varepsilon$ in the Randall-Sundrum model with SM bulk gauge group~\cite{Bauer:2009cf}. The different colours correspond to different choices for the hadronic parameters.\label{fig:eps}}
\end{figure}

The review of possible NP effects in rare $K$ decays would be incomplete without briefly discussing the status of the CP violating parameter $\varepsilon'/\varepsilon$. Measuring the ratio of direct versus indirect CP violation in $K_L\to\pi\pi$ decays, it is sensitive to the same NP contributions that yield large enhancements of the CP violating decays $K_L\to\pi^0\nu\bar\nu$ and $K_L\to\pi^0\ell^+\ell^-$. This ratio has been measured rather precisely~\cite{Beringer:1900zz}
\begin{equation}
\text{Re}(\varepsilon'/\varepsilon)_\text{exp} = (1.66\pm0.23)\cdot 10^{-3}
\end{equation}
and can in principle place strong bounds on the possible NP effects in $K_L\to\pi^0\nu\bar\nu$ and $K_L\to\pi^0\ell^+\ell^-$. The correlation between $\varepsilon'/\varepsilon$ has been studied explicitly in a number of models, including supersymmetry~\cite{Buras:1999da}, the Littlest Higgs model with T-parity~\cite{Blanke:2007wr}, the Randall-Sundrum model with SM bulk gauge symmetry~\cite{Bauer:2009cf}, the SM with a sequential fourth generation~\cite{Buras:2010pi} and the modified electroweak penguin scenario~\cite{Buras:2004ub}. The common outcome of all these studies is that in order to derive stringent constraints from the measured value of $\varepsilon'/\varepsilon$ a much better understanding of the relevant hadronic matrix elements is needed. As an example Fig.\ \ref{fig:eps} shows the correlation between $K_L\to\pi^0 \nu\bar\nu$ and $\varepsilon'/\varepsilon$ in the Randall-Sundrum model with SM bulk gauge group~\cite{Bauer:2009cf}, where different colours correspond to different choices for the hadronic parameters. Progress in the evaluation of the relevant matrix elements is therefore badly needed. For an overview of the present status of lattice calculations see~\cite{Christ}.

\section{\boldmath $K\to\ell\nu$ and lepton non-universality}

While neutral kaon mixing and the rare and CP violating $K$ decays discussed so far are sensitive to new flavour and CP violating interactions in the quark sector, also the lepton flavour sector can be probed through kaon physics. The ratio
\begin{equation}
R_K= \frac{\Gamma(K\to e\nu)}{\Gamma(K\to\mu\nu)}
\end{equation}
measures the amount of lepton non-universality between the first and second generation and is theoretically very clean. The recent data~\cite{Lazzeroni:2012cx}
\begin{equation}
R_K^\text{exp} = (2.488\pm0.010)\cdot 10^{-5}
\end{equation}
are in good agreement with the SM prediction~\cite{Cirigliano:2007ga}
\begin{equation}
R_K^\text{SM} = (2.477\pm0.001)\cdot 10^{-5}\,,
\end{equation}
albeit with still an order of magnitude larger uncertainties. An improved measurement will therefore yield a significant constraint on NP.

\begin{figure}
\center{\includegraphics[width=.4\textwidth]{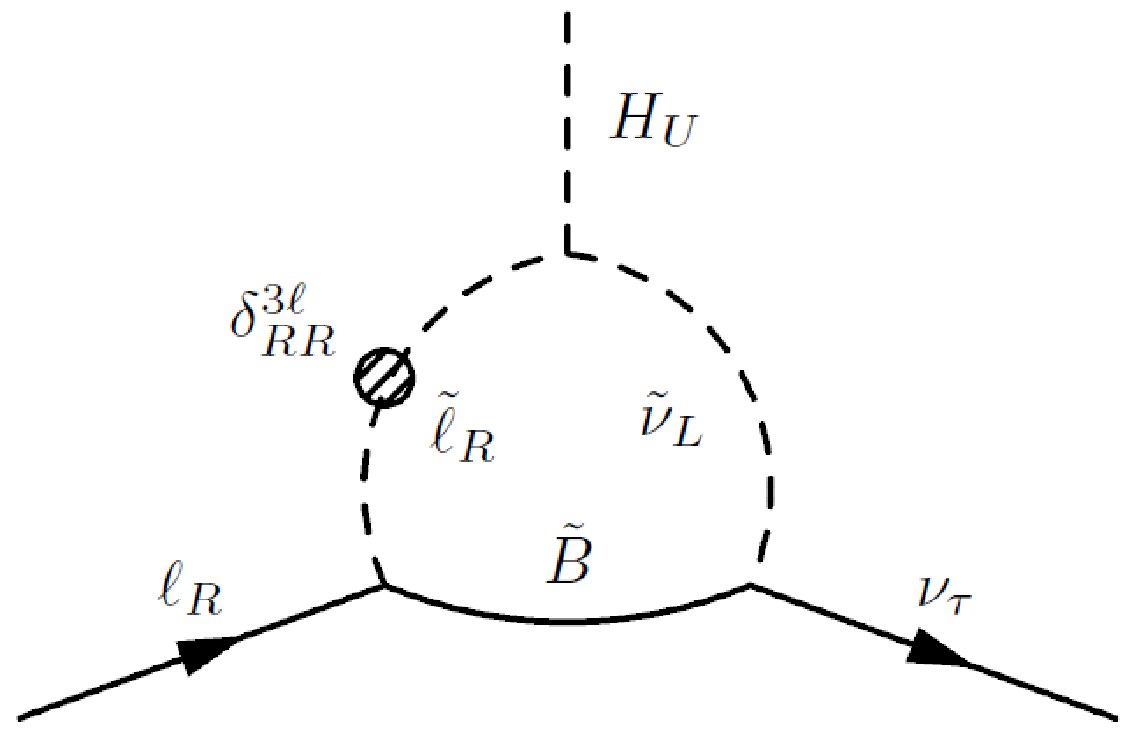}}
\caption{Lepton non-universality in the MSSM generated through lepton flavour violating interactions~\cite{Masiero:2008cb}.\label{fig:Kl2}}
\end{figure}

A deviation of $R_K$ from the SM prediction could for instance be generated by lepton flavour violating interactions present in the MSSM, as depicted in Fig.\ \ref{fig:Kl2}~\cite{Masiero:2008cb}. The test of lepton universality in $K\to\ell\nu$ decays therefore provides a complementary test of lepton flavour violation. It is interesting to note that in this scenario there is no interference between SM and NP contributions, so that $R_K$ is always enhanced over its SM value. 

Last but not least let us note that a deviation from the SM prediction of $R_K$ could also originate from the presence of sterile neutrinos~\cite{Shrock:1980ct}.

\section{Summary}

Kaon physics offers a particularly powerful probe of non-SM flavour violating interactions.  
Due to their theoretical cleanness and strong suppression in the SM  the rare $K\to\pi\nu\bar\nu$ decays constitute a unique possibility to identify new contributions even from particles in the multi-TeV range. Large deviations from the SM are in fact possible in a wide range of popular NP scenario, despite the non-observation of NP in direct searches and very SM-like signatures at the LHCb. Furthermore the correlation between the charged and neutral decay rate allow to shed light on the operator structure in $K^0-\bar K^0$ mixing, constrained most stringently by $\varepsilon_K$, and its correlation with $\Delta S=1$ flavour violating transitions. A precise measurement of both the $K^+\to\pi^+\nu\bar\nu$ and $K_L\to\pi^0\nu\bar\nu$ branching ratios is therefore of utmost importance.

Additional information on the NP operators contributing to rare $K$ decays can be obtained from the study of the $K_L\to\pi^0\ell^+\ell^-$ decays. While theoretically less clean than the $K\to\pi\nu\bar\nu$ modes, these channels can be affected by scalar operators and thus their correlation offers a powerful tool to distinguish between various NP scenarios. 

An important constraint on CP violation in rare $K$ decays will be placed by the parameter $\varepsilon'/\varepsilon$, once a more precise SM prediction is available from the lattice.

Finally, the ratio of $K\to\ell\nu$ decays yields a sensitive probe of lepton non-universality beyond the SM and is complementary to the searches for lepton flavour violating interactions in rare $\mu$ and $\tau$ decays.

\end{document}